# Morphing a Stereogram into Hologram

Enrique Canessa[1] and Livio Tenze
*Science Dissemination Unit (SDU)*
*ICTP - International Centre for Theoretical Physics, Trieste, Italy*

**Abstract**

This paper develops a simple and fast method to reconstruct reality from stereoscopic images. We bring together ideas from robust optical flow techniques, morphing deformations and lightfield 3D rendering in order to create unsupervised multiview images of a scene. The reconstruction algorithm provides a good visualization of the virtual 3D imagery behind stereograms upon display on a headset-free Looking Glass 3D monitor. We discuss the possibility of applying the method for live 3D streaming optimized via an associated lookup table.

**Introduction**

Virtual reality (VR) to engage with 3D objects and environments has been a burgeoning and popular subject in recent years. This technology is highly beneficial in a range of diverse innovative scientific, engineering and entertainment applications including medicine, astro stereo photography, robotic vision, intelligent transportation systems and video games among others [1]. For example, it has been recently shown that the advantages of 3D stereoscopic visualization over a conventional 2D planar screen can shorten the dissection time of specialized surgery [2]. Although many advances have been achieved already, a key limitation for VR technologies is the fact that, in some cases, it can produce contagious yawning [3]. It is also necessary to wear special goggles or headsets, causing motion sickness and limiting user time. Notwithstanding such constraints, VR applications can still give science a new dimension allowing researchers to view and share 3D data [4]. The final goal however would be to obtain and visualize similar results without wearing any device.

Motivated by the study of Ryan Baumann on animating stereograms with optical flow morphing [5], in this paper we made a first attempt to reconstruct reality in similar simple terms. We develop a method to combine observations of 2D stereoscopic images with 3D virtual interpretations of reality. We apply the unsupervised *torch-warp* optical flow algorithm [5,6] to animate stereo pairs and retrieve distinct 2D views from morphing deformations between left (L) and right (R) images.

Since all these sequential frames put together can give an acceptable illusion of depth and parallax in the horizontal direction [5], we place them in a single standard Quilt collage [7]. We then convert the Quilt into a native lightfield image using open source SURF*sara* visualization python scripts [8]. This Quilt provides a good visualization of virtual 3D imagery of stereograms by a direct display of the lightfield output images on the new class of standalone Looking Glass 3D monitors [9]. Multiple viewers can see the scene inside stereoscopic images without the need for glasses and from different angles. We discuss on the possibility of applying this method for producing multiview 3D streaming with just a stereo webcam.

---
[1] Author for correspondence: canessae@ictp.it

**Hardware**

We use low-cost ELP-960P2CAM (V90 and LC1100) USB stereo webcams with no distortion dual lens and M12 mount synchronization to obtain the principal 2D stereoscopic images [10]. According to specifications, the two camera video –with low power consumption, 90 degree lens, standard electronic rolling shutter and 1/3" CMOS OV9750 sensor for high quality image– can reach high frame rates in MJPEG compression format of 2560(H)x960(V)p@60fps with a sensitivity 3.7V/lux-sec@550nm. Its small size of 80x16.5 mm is useful for embedded applications. It supports Linux OS USB video class UVC with adjustable parameters such as brightness, contrast, saturation, hue, sharpness, color balance and exposure.

For the multiple angles visualization of our morphing to hologram image files generated in full color with the present algorithm, we use the glasses-free standard Looking Glass 8.9″ (also known as HoloPlay) as an HDMI external monitor [9]. This HoloPlay device combines lightfield and volumetric technologies within a single new type of display [10], which allows to display an hologram of simultaneous 32 (or 45) different views at 60 fps formed via a 8x4 (or 9x5) Quilt input. The technology used in this class of monitor is described in the U.S. Patent application number 2017-0078655 : *"Printed Plane 3D Volumetric Display"*.

Besides the standard 64bit Windows 10, the viewing under O.S. Linux Ubuntu 19.04 was also possible using a Notebook Aspire E 15, Inter Core i5, 64bit, 8GIB RAM, 1366x768p resolution with graphics card Nvidia GeoForce 820M output 2560x1600p. The small Raspberry Pi 3 single-board computer device Model B+ 1.4GHz 64-bit quad-core processor with extra power supply was also able to display our stereograms morphed into multiview 3D display.

**Method**

The procedure adopted to get a stereogram is as follows: we first position the principal object in the scene, at least 1.6 meter distant apart from the two lenses of the ELP stereo webcam. This distance enables us to avoid image distortions due to well-known technical limitations of stereoscopic webcams [11]. The dual lenses are aligned parallel toward the horizon, avoiding to tilt the ELP's device.

The most outstanding feature of the compact ELP synchronized stereo webcam we use is that the two cameras video frames are synchronous. This unique feature enables to simulate the manner in which human eyes observe 'simultaneously' one scene from two different viewpoints [10]. This is ideal for binocular stereo vision development like the one studied in this work. By one single shot we retrieve a single image in HD resolution, containing L and R views, and without the need for any extra, complex prior calibration of the stereo webcams as in the case of most compact industrial camera devices producing and displaying two L and R images independently.

We take with the ELP a stereoscopic test picture using the *ffmpeg* command and cut the resulting single HD image of max. 1280x960p resolution in two equal parts to generate the L and R images. We then resize each view to set initially the widths to 512p and then crop their heights to 256p (anywhere along its vertical axis starting from the same upper corner in both images). This image manipulation is carried out before the morphing rendering to speed up the computation using smaller images and avoid the case of an '*out-of-memory (RAM) condition before sgemm*' in the deep convolution matching [6] when trying to match the full HD images.

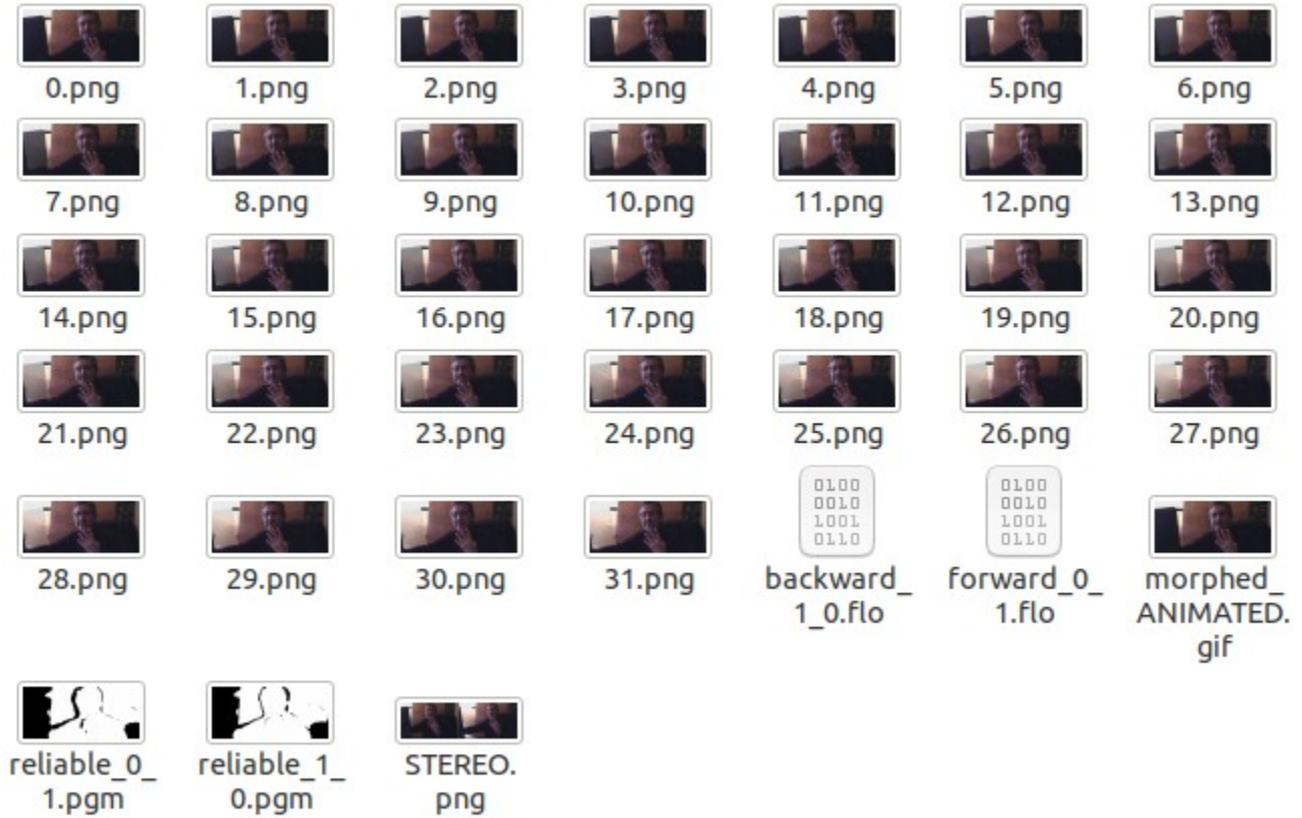

**Fig. 1**: *32 intermediate views (0-31 frames) by morphing a stereoscopic image via the 'torch-warp' algorithm described in [5]. Shown are also the morphed animated Gif and the backward and forward displacements of the optical flow data.*

Our method for 3D multiview then follows closely the algorithm of Baumann for animating stereograms [5]. We automatically apply optical flow deformations based on DeepFlow (which outputs *.flo* files) to our pairs of L-R images in order to morph between them. This alignment of the two (dissimilar) images, followed by a gradual fade out from one image to the other, can give an acceptable illusion of depth and (parallax) motion to the viewer (see examples in [5]).

In particular, this continuous interpolation by optical flow-based image warp enabled us to control the gradually cross-dissolving between pairs of stereo images. By fading out from one L image to the other R, we can then split a whole scene into a given number of intermediary component frames as illustrated in Figure 1.

We first generate .flow data for 32 different views (frames numbered 0-31 in the figure) starting from L to R images in an entirely unsupervised manner. This choice for the morphing process is done based on the nature of the Looking Glass HoloPlay [7].

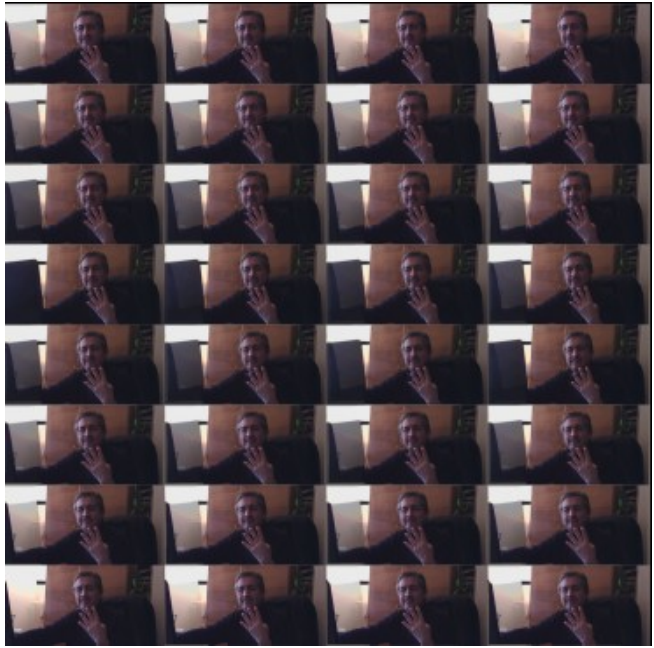

**Fig. 2**: *Example of 8x4 Quilt input for the Looking Glass HoloPlay from the intermediary component (morphed) frames in Figure 1.*

Next, the 32 different (*.png*) views from the morphing data are then converted to PNG24 to form the needed Quilt. We place the set of 32 view images (512x256p) in a single standard 4x8 Quilt image (2048x2048 p) as in Figure 2 using *'make_quilt.py'* by SURFsara scripts [8]. The views start from the bottom left as the leftmost view (L-image) to the top right being the furthest right (R-image).

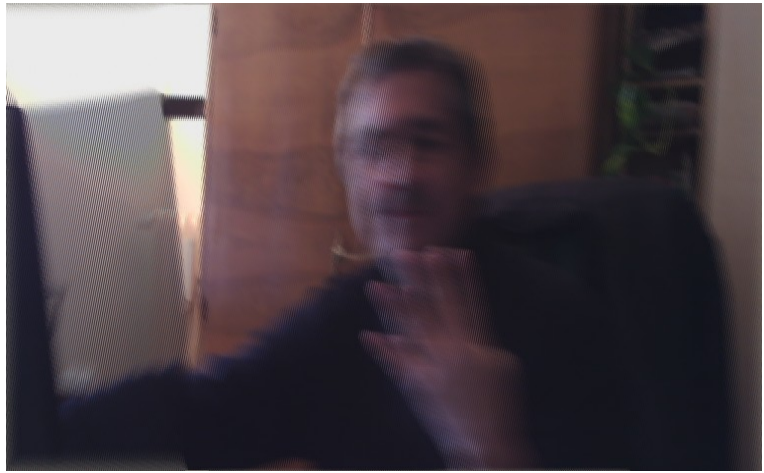

**Fig. 3**: *Native image output of the Quilt in Figure 2 based on the per-device HoloPlay calibration '.json' values.*

Finally, we generate a native image targeted to a specific HoloPlay device using the *'quilt2native.py'* script in [8] as shown in the example of Figure 3. We get the display calibration values (in the form of a standard data interchange file *.json*) from a Looking Glass Display using: *'get_calibration_from_eeprom.py'* from [8].

The multiview 3D rendering of our morphed stereogram of Figure 3 can then be displayed on a HoloPlay device as appears in Figure 4. This output can also be seen in the video: https://www.youtube.com/watch?v=6FAhmI-vtLQ

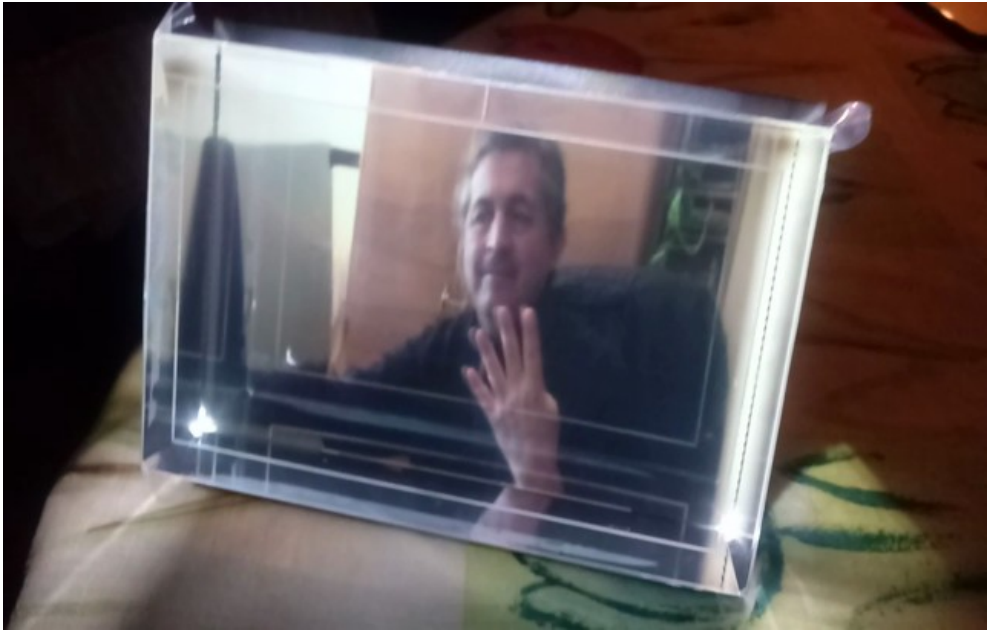

**Fig. 4**: *Multiview hologram output from morphed stereoscopic images. See also video demo at: https://www.youtube.com/watch?v=6FAhmI-vtLQ*

In order to produce beautiful holograms, as in Figure 4, Looking Glass provides 32 (or 45) discrete views or frames of a 3D scene, displaying these views over a ~50°-wide view cone. This lightfield arrangement tricks our visual perception into seeing 3D objects by parallax (*i.e*, moving the head around the scene, and by stereo vision (*i.e.,* presenting different perspectives to each eye*).*

**Discussion**

Each Looking Glass holds own calibration data for correct rendering and, inside its render volume, different depths have different optical properties [7]. The depth where things look sharpest is at the so-called Zero-Parallax Plane in the middle of the display (in our case, around frame Nr. 16) . Objects in this plane show up in the same pixel-space position for all 32 (or 45) views. Objects in the scene that are nearer or further than this plane, undergo parallax. The Looking Glass HoloPlay provides a novel glasses-free way to preview 3D objects and scenes as in Figure 4.

A generic expression for the relation between the pixels of a slanted lenticular 3D-LCD and the multiple perspective views was first derived by Cees van Berkel [12]. Each sub-pixel on the 3D-LCD is mapped to a certain view number and colour value (*i.e.,* in the lightfield domain). If *i* and *j* denote the panel coordinates for each sub-pixel, then

$$N_{i,j} = N_{tot}\ (i - i_{off} - 3j\ tan(\alpha))\ mod(P_x)/P \quad , \qquad (1)$$

where *N* denotes the view number of a certain viewpoint, *α* the slanted angle between the lenticular lens and the LCD panel and $P_x$ the lenticular pitch. Upscaling multiple views (*e.g.,* upscaling from the Quilt to the native Looking Glass image) requires lots of CPU resources and increases system complexity [13-15].

We have created multiview images via Eq.(1) starting from a stereoscopic scene. We have put together ideas from optical flow techniques, morphing deformations and lightfield 3D rendering. 2D morphing yield reasonable and better 3D visual results when it works correctly along the horizontal direction, as required by the optical elements adopted by the Looking Glass HoloPlay [7]. Fading from misaligned L to R images can cause the other intermediate parts of the whole scene to blur and get distorted. The limits of the morphing process in stereo animation, as compared with the use of depth map, is that morphing cannot provide complete information on distant regions. It mainly extrapolates the parallax encoded in the images by combining information from nearby pixels only. It is essentially a purely local method [16].

However, the application of 2D morphing to create the (32 or 45) required 2D views for the Looking Glass Quilt poses minimum geometric constraints on the reconstructed 3D scene via the alternate lightfield projections. Morphing between the two images taken simultaneously with the ELP stereo webcam produces a nice illusion of 3D throughout the multiviews of a scene.

Optical flow techniques are relatively sensitive to the presence of occlusions, illumination changes and out-of-plane movements. These factors lead to noise and to obtain translation motion discontinuities between the neighborhoods of two consecutive images. The key processing steps in the flow field [5,6], include the matching by polynomial interpolation to approximate pixel intensities in the neighborhood, warping and optimization without an explicit regularization. We have found that to generate .flow data for 32 views, optical flow techniques can lead to a reasonable accuracy to reconstruct 3D reality from stereoscopic images. The CPU time for creating these sets of view frames, and especially the final Looking Glass native image, can take a few minutes. This process still needs to be optimized.

To map the Quilt into the final native image for a display on the holographic Looking Glass HoloPlay, it is necessary to apply the complex algorithm of Eq.(1) that depends on input parameters of the physical structure of this device. Every HoloPlay monitor in fact possesses unique calibration parameters (such as pitch, slope, dpi) set in the phase of manufacturing. By making use of this calibration, and applying lightfield geometric transformations, one then gets the multiview image reproduced in the Looking Glass HoloPlay. This mapping procedure requires considerable calculation power, since the final native image is at a resolution of 2560x1600p with 3 color channels –such that, once the pixel to be mapped is fixed, the map value for each color channel implies separated calculations. In essence this procedure, as such, would become computationally expensive and difficult to apply in applications for a real-time video in 3D.

Since the mapping matrix depends on the geometric position of each pixel, and on the calibration parameters of the display, we have constructed a Lookup Table (LUT) to replace runtime computation and save processing time. This LUT is created only once at the beginning of the mapping process: Quilt→HoloPlay image, and then used for all the images that need to be visualized. We create the array as follows: First, we allocate 3 matrices for the three color channels RGB of size 2560x1600p x 2. Each matrix provides the X coordinate of the Quilt from which we take the corresponding value and the Y coordinate. This explains the multiplication of the resolution 2560x1600p by 2. To avoid unnecessary waste of resources and consume the least possible amount of RAM memory, each element of the matrices is made of type uint16_t (the uint8_t type would allow to address maximum values of 255). Secondly, all the positions of the pixels 2560x1600p are scrolled and we calculate the mapping value for each pixel on the Quilt image. Next, once the mapping value has been calculated, the value is stored in the three different allocated matrices.

Finally, once the map filling procedure has been completed, we save the 3 three matrices in binary format. Then, at each successive time step it is possible to reload the matrices (without recalculating them) and apply the mapping automatically to all the necessary images. This procedure allows to speed up significantly the mapping procedure –the rendering operation of the final native image is essentially achieved by accessing the elements of the 3 matrices to map the Quilt pixels on the final native image for a dynamical display on the HoloPlay.

We have implemented anew a C-library that includes the LUT to quantify by benchmarking the timing to convert the Quilt into the native image between the direct, classical method of Eq.(1) –or, similarly to those of the SURFsara scripts [8]. A test with *gprof f*or the statistics of the single files inside the C-library indicates that the total conversion time is 1.18s against 0.67s for the LUT.

This means that the implementation of LUT allows to reduce the computing time of about 50%. One could even reduce the computation time by a factor of 4, by allocating the 4 threads of a Raspberry Pi 3 with 4 processors to the Quilt image of Figure 2 divided in four parts (with successive 8 views each). These possibilities open the path for producing multiview 3D streaming in real-time with a simpler and faster algorithm.

**Conclusions**

We have proposed a rapid conversion approach that allows to transform stereoscopic images into holograms via the automatic implementation of a LUT and unsupervised morphing deformations between the L and R images and with no use of depth map. The key idea here is just to start from the pair of structured stereoscopic 2D images with no additional information about any intermediate view to create the holograms. Under this method, we establish correlations between 2D image observations and their 3D display on a Looking Glass HoloPlay monitor.

Future work could include the potential application of this method in the field of live 3D streaming using cost effective hardware and no headsets. In principle it would be also possible to interact with such real-time holograms displayed in the HoloPlay screen with a Leap Motion controller [7].

Ultimately, the different set of techniques discussed in this work still need to be optimized and further experimented when combined all together. Work along these lines is under development.